\newcommand{\bq}{\begin{equation}}
\newcommand{\eq}{\end{equation}}
\newcommand{\bqa}{\begin{eqnarray}}
\newcommand{\eqa}{\end{eqnarray}}

\newcommand{\nl}{\nonumber \\}
\newcommand{\f}{\varphi}
\newcommand{\al}{\alpha}
\newcommand{\ddv}[1]{{\partial\over\partial #1}}
\newcommand{\ddvv}[2]{{\partial^{#1}\over(\partial #2)^{#1}}}
\newcommand{\vvf}{\vec{\f}}
\newcommand{\intl}{\int\limits}
\newcommand{\suml}{\sum\limits}

\documentclass[a4paper]{article}
\usepackage{verbatim}
\usepackage{moreverb}
\usepackage{latexsym}
\usepackage{axodraw}
\usepackage{graphicx}
\usepackage{epsfig}

\title{Non-Concave Bare Actions with Concave Effective Actions I: Zero Dimensions}
\author{E.N. Argyres\footnote{Institute of Nuclear Physics, NCSR "Demokritos", Athens, Greece} \and M.T.M. van Kessel\footnote{Radboud Universiteit Nijmegen, Nijmegen, The Netherlands, M.vanKessel@science.ru.nl} \and R.H.P. Kleiss\footnote{Radboud Universiteit Nijmegen, Nijmegen, The Netherlands, R.Kleiss@science.ru.nl}}

\begin{document}

\maketitle

\section{Abstract}

We perform an explicit computation of the effective action for a few zero-dimensional Euclidean field theories in which the bare action exhibits several (or infinitely many) minima. In all cases the effective action is well-defined for all field values, as well as purely concave, in contrast to the bare action. We give arguments against the validity of the naive perturbative approach, which would lead to spontaneous symmetry breaking.

\section{Introduction}

In this article we will consider Euclidean field theories in which the bare action has two or more minima. For now we will stick to field theories in zero dimensions. Field theories in higher dimensions will be considered later \cite{NextPaper}.

In the case of zero-dimensional Euclidean field theories it is easily proven that the effective action has to be concave. Consider such a field theory with $n$ fields $\f_i$, $\vvf = (\f_1, \f_2, \ldots, \f_n)$, and $n$ sources $J_i$, $\vec{J} = (J_1, J_2, \ldots, J_n)$. Denote the generating function of the connected amplitudes by $\vec{\phi}(\vec{J})$. The effective action will be denoted by $\Gamma(\vec{\phi})$. The definition of concavity is that
\bq
\frac{\partial^2}{\partial \alpha^2} \Gamma(\vec{\phi}+\alpha\vec{\rho}) \Big|_{\alpha=0} > 0
\eq
for all $\vec{\phi}$ and $\vec{\rho}$. This means that
\bq
\rho_i \rho_j \frac{\partial^2\Gamma}{\partial \phi_i \partial \phi_j}(\vec{\phi}) > 0.
\eq
Expanding the vector $\rho$ in terms of eigenvectors of $\frac{\partial^2\Gamma}{\partial \phi_i \partial \phi_j}(\vec{\phi})$:
\bq
\vec{\rho} = \sum_j c_j \vec{e}_j,
\eq
we get
\bq
\sum_j c_j^2 \lambda_j > 0,
\eq
where the $\lambda$'s are the eigenvalues. Now this should hold for all $c$'s, thus all eigenvalues $\lambda$ are greater than zero for all $\vec{\phi}$. This argument can also be inverted to show that if all eigenvalues are greater than zero, the effective action is concave.

Now for a Euclidean field theory we obviously have that
\bq
\rho_i \rho_j \langle \f_i \f_j \rangle_{\vec{J}} = \langle (\rho \cdot \f)^2 \rangle_{\vec{J}} > 0
\eq
where $\langle \rangle_{\vec{J}}$ is the path integral average in the presence of the source $\vec{J}$. Note that the result $\langle (\rho \cdot \f)^2 \rangle_{\vec{J}} = 0$ is conceivably possible but only if the path integral measure contains weird constraints, which case we will not consider. This means that the matrix $\langle \f_i \f_j \rangle_{\vec{J}}$ only has eigenvalues greater than zero. Now the effective action satisfies
\bq
\frac{\partial^2\Gamma}{\partial \phi_i \partial \phi_j}(\vec{\phi}) = \hbar \langle \f_i \f_j \rangle_{\vec{J}(\vec{\phi})}^{-1} \;\;,
\eq
which means that the eigenvalues of $\frac{\partial^2\Gamma}{\partial \phi_i \partial \phi_j}(\vec{\phi})$ are also greater than zero. Thus for a zero-dimensional theory with $n$ fields the effective action is concave.

In the standard treatment of Euclidean field theories where the bare action has more than one minimum there is a problem with the effective potential. This effective potential is not concave and becomes complex or undefined in certain regions. This phenomenon can easily be understood as follows. The bare action can only be a good approximation to the effective action if it shares with it the fundamental property of concavity. If, therefore, we start to expand the theory around some minimum of the bare action, this expansion will only be appropriate in at most that region around the minimum where the bare action is concave. Where concavity breaks down, there so does the perturbative expansion.

In the standard treatment there is also another problem. In this standard treatment one chooses one of the minima of the bare action and expands around it. The path integral approach to field theory, however, suggests that all the minima of the bare action should be included in the treatment.

This last problem will appear to be closely related to the first one. Our treatment, presented for a few theories in this article, will use the path integral approach and include all minima accordingly. We will show that this solves the first problem, the effective potentials we get will be concave and well-defined. As a secondary result of this new treatment we will not get spontaneous symmetry breakdown, in contrast to the standard treatment. 

This all was already pointed out in \cite{BenderCooper} and \cite{CooperFreedman} in the case of $\varphi^4$-theory with two minima. In this article we will however also treat field theories with two fields and an infinite set of minima.

\section{The zero-dimensional $Z_2$-model}

\subsection{Equations for the effective action}
We consider a zero-dimensional theory with a bare action given by 
\bq
S(\f) = {\lambda\over4!}\f^4 - {\mu\over2!}\f^2 - J\f.
\eq
Here $\f$ is the single, real-valued, dynamical field variable, and $J$ is the source; $\lambda,\mu>0$ so that this is a well-defined theory with two minima, given by the equation
\bq
\ddv{\f}S(\f) = {\lambda\over6}\f^3 - \mu\f - J  = 0.
\eq
For $J=0$, the action has a local maximum at $\f=0$ and two equally deep
minima at $\f=\pm v$, where 
\bq
v = \sqrt{{6\mu\over\lambda}}.
\eq
As $|J|$ increases from zero, the degeneracy is lifted and the positions of the saddle points shift on the real axis, until also $\partial^2S(\f)/(\partial\f)^2 = 0$ at one of the saddle points, which happens for $J=\pm 2v\mu/(3\sqrt{3})$, $\f=\mp v/\sqrt{3}$, with the remaining minimum of the action at $\f=\pm2v/\sqrt{3}$. For stronger sources, there is only one minimum of the action situated on the real axis.

The zero-dimensional path integral is given by
\bq
Z = \int\;e^{-S(\f)/\hbar}\;d\f.
\eq
It obeys a Schwinger-Dyson equation,
\bq
\left({\lambda\over6}\hbar^3\ddvv{3}{J} - \mu\hbar\ddv{J} - J\right)Z(J) = 0,
\eq
as well as a `stepping equation' (see \cite{Argyres} and \cite{Cvitanovic}):
\bq
\left(\ddv{\mu}-{\hbar\over2}\ddvv{2}{J}\right)Z(J) = 0.
\eq
Notice that the stepping equation holds for any action that contains the term $-\mu\f^2/2$.

The generating function of the connected amplitudes is given by
\bq
\phi(J) = \hbar\ddv{J}\ln(Z);
\eq
under this definition, the $n$-th derivative of $\phi(J)$ to $J$ at $J=0$ gives the $(n+1)$-point connected amplitude. In terms of $\phi(J)$, the Schwinger-Dyson and stepping equations read
\bq
{\lambda\over6}\left(\phi^3 + 3\hbar\phi\ddv{J}\phi + \hbar^2\ddvv{2}{J}\phi \right) - \mu\phi - J = 0,
\eq
and
\bq
\ddv{\mu}\phi - \phi\ddv{J}\phi - {\hbar\over2}\ddvv{2}{J}\phi = 0,
\eq
respectively.

The effective action, $\Gamma(\phi)$ is by definition a function of $\phi$ such that the following implicit equation holds:
\bq
J(\phi) = \ddv{\phi}\Gamma(\phi).
\eq
We can therefore write the Schwinger-Dyson and stepping equations in terms of $\Gamma$ rather than in terms of $\phi$. For ease of discussion, we shall employ a dimensionless formulation, by writing
\bq \label{dimlessvars}
\phi = vf, \quad J=v\mu x, \quad \hbar = g\mu v^2,
\eq
and
\bq
\ddv{\phi}\Gamma(\phi) = v\mu F(f,g).
\eq
Note that since the effective action, like the bare action, is only defined up to a constant additive term, the function $F$ contains all relevant information about the physics of the theory. The  Schwinger-Dyson and stepping equations for $F$ now read
\bq
f^3 - f - F + 3gf\left(\ddv{f}F\right)^{-1} - g^2\left(\ddvv{2}{f}F\right)\left(\ddv{f}F\right)^{-3} = 0
\eq
and
\bq
f + {3\over2}F - {1\over2}f\left(\ddv{f}F\right) - {1\over2}g\left(\ddvv{2}{f}F\right)\left(\ddv{f}F\right)^{-2} - 2g\left(\ddv{g}F\right) = 0,
\eq
respectively. The second-derivative term may be eliminated to yield
\bq
(f^3-f-F)\left(\ddv{f}F\right) + g\left(f\left(\ddv{f}F\right) + f -3F\right) + 4g^2\left(\ddv{g}F\right) = 0. \label{basic}
\eq
We shall use equation (\ref{basic}) to determine various approximations to the effective action. An additional condition on the solution is provided by the fact that the effective action must, like the bare one, be symmetrical under $\phi\to-\phi$, from which we have
\bq
F(0,g) = 0.
\eq

\subsection{Flat and steep solutions}
By imposing a perturbative expansion on $F$:
\bq
F(f,g) = \suml_{n\ge0} g^nF_n(f),
\eq
we can transform equation (\ref{basic}) into a hierarchy of equations for the $F_n$. The lowest-order equation reads, of course:
\bq
(f^3-f-F_0)\left(\ddv{f}F_0\right) = 0.
\label{lowest}
\eq
The first possible solution of equation (\ref{lowest}) is
\bq
F_0 = f^3 - f,
\eq
which corresponds to the procedure in which {\bf one} of the two minima of the action is used as the starting point for a straightforward perturbative expansion. The hierarchy of equations is now of algebraic type, in the sense that the $n$-th order equation contains $F_n$ but no derivative of $F_n$, and all derivatives are of lower-order $F$'s. We arrive at the `steep' solution
\bq
F(f,g) = f^3 - f + g{3f\over3f^2-1} + g^2{3f(3f^2+5)\over(3f^2-1)^4} + g^3{180f(6f^2+1)\over(3f^2-1)^7} + {\cal O}(g^4).
\eq
It is easily checked that this `steep' solution satifies both the Schwinger-Dyson and stepping equation. Attractive as it may seem, this solution suffers from a serious drawback. Beyond the lowest order, the effective action is undefined for $f^2=1/3$. Even if we accept this as a fact of life, the problem remains that $F(f,g)$ is {\bf not monotonic in any order of perturbation theory} even for $f^2>1/3$, in contradiction with the general result that {\bf the effective action must be concave}, at least for a Euclidean theory with real $\f, J$ and real action. We regard this as the signal that this perturbative approach is actually invalid.

A point to be made here is the following. As we have seen, the two minima of the bare action move about the real axis as $J$ increases: for the critical value $x=2/\sqrt{3}$, the minimum at $\f=-v/\sqrt{3}$  becomes a point of inflexion and is no longer a valid starting point for {\bf any} perturbative treatment. However, that does not mean that $\phi$, the generating function, obeys $f^2=1/3$ there, as we shall see, in fact we have $f>1$ at that point. The failure of perturbation theory at $f^2=1/3$ is therefore not so simply related to the disappearance of a minimum of the bare action.

There is, however, the alternative solution to equation (\ref{lowest}):
\bq
\ddv{f}F_0 = 0\;\;\;\Rightarrow\;\;\;F_0 = 0,
\eq
where we have used the symmetry of the effective action. The perturbative series hence starts at order $g^1$ rather than at order $g^0$, leading to an effective action that is of order $\hbar$ rather than of order $\mu v^2$, hence the name `flat'. The lowest nontrivial solution, $F_1$, is now given by 
\bq
F_1 = {1\over2}L, \quad L=\log\left({1+f\over1-f}\right),
\eq
and including some higher orders we find
\bqa
F(f,g) &=& g{L\over2} + g^2{(1-3f^2)L-6f\over8(1-f^2)} + \nl 
& & g^3{(1-f^2)^2L^3 + (10f-6f^3)L^2 + (17-138f^2+105f^4)L+222f-210f^3\over64(1-f^2)^2} + \nl
& & \mathcal{O}(g^4).
\eqa
This solution, too, obeys the Schwinger-Dyson and stepping equations. It is in the `flat' solution that that stepping equation becomes important. If we would only use the Schwinger-Dyson equation (which  by itself is already sufficient to obtain the `steep' solution, as can easily be checked), the differential equations for the $F_n$ would be of second order. The symmetry of the effective action provides one condition, but this only reduces the solution to a one-parameter family of (a priori) equally acceptable solutions left. Apparently, the stepping equation recovers the additional needed piece of information.

The `flat' solution is defined as long as $f^2<1$. For larger values of $f$, where the steep solution may be considered as a better approximation to the real physics of the theory, the leading term of $F$ is, as we have seen, of order 1 rather than of order $g$. The divergence at $f^2\approx1$ is therefore not surprising. However, since the regions of apparent validity of the flat and steep solutions overlap, it behooves us to investigate the switch from one to the other. This we shall do in the  next section.

\subsection{Saddle-point approximation}
Let us assume that $|x|<2/3\sqrt{3}$ so that the bare action contains two minima. These are given by
\bq
f^3 - f - x = 0.
\eq
The exact solutions to this equation can conveniently be parametrized as follows:
\bq
x = {2\over3\sqrt{3}}\sin(3\al), \quad f_\pm = {2\over\sqrt{3}}\sin\left(\al\pm{\pi\over3}\right)
\eq
where $-\pi/6<\al<\pi/6$. We may also introduce
\bq
S_\pm = {1\over\hbar} S(\f_\pm) = -{1\over3g}\sin\left(\al \pm {\pi\over3}\right)\left( \sin\left(\al \pm {\pi\over3}\right) + \sin(3\al)\right)
\eq
and
\bq
\Sigma_\pm = {1\over\mu}\ddvv{2}{\f}S(\f_\pm) = 4\sin\left(\al \pm {\pi\over3}\right)^2 - 1.
\eq
The subleading minimum gives a nonperturbative contribution, but `nonperturbative' is not the same as `negligible', as we shall see. The saddle-point approximation to $Z$ and to $\phi$ is now given by
\bqa
Z &\approx& Z_+ + Z_-, \nl
f &\approx& {1\over Z}\left( f_+\left(1-{3g\over\Sigma_+^2}\right)Z_+ + f_-\left(1-{3g\over\Sigma_-^2}\right)Z_-\right), \nl
Z_\pm &=& \Sigma_\pm^{-1/2}\exp(-S_\pm).
\eqa
We may trust this solution as an approximation to the physics as long as $\Sigma_\pm^2$ is appreciably larger than $3g$. Since both $f$ and $x$ are determined by the parameter $\al$, it is easy to plot their relation. Below, we give the result, for $g=0.03$, for the (derivative of) the effective action in various approximations.
\begin{center}
\includegraphics[width=8cm, angle=-90]{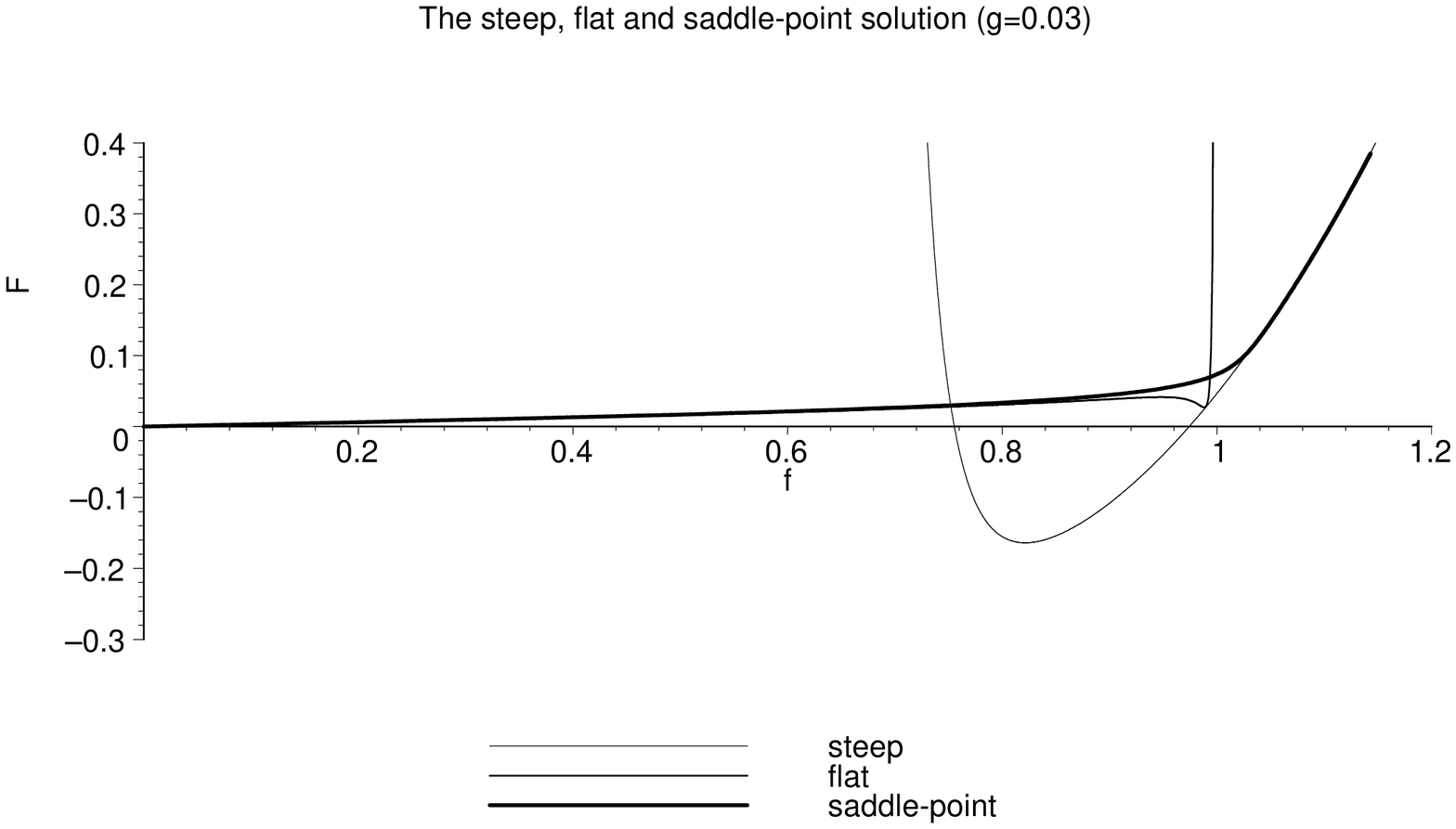}
\end{center}

The flat solution is a good approximation for small $f$, whereas the steep solution is a good approximation for large $f$. The saddle-point solution interpolates between the two and is a good approximation to the true solution everywhere. Indeed we find this solution to be concave and well-defined everywhere.

\section{The zero-dimensional $O(N)$-model}

\subsection{A critique of symmetry breaking}

We may extend the previous result to the case of $N$ real-valued fields $\f_1$, $\f_2$, $\ldots$, $\f_N$. The action without the source term is given by
\bq
S(\vvf) = {\lambda\over24}(\vvf^2)^2 - {1\over2}\mu\vvf^2 + {3\over2} {\mu^2\over\lambda}.
\eq
In the absence of sources, the minima of the action lie on the surface $\vvf^2=v^2\equiv{6\mu\over\lambda}$. Taking an arbitrary point on this surface as the `real' minimum of the action leads to problems that are even worse than in nonzero dimensions, since the resulting $N-1$ Goldstone modes corresponding to the flat directions have, in the absence of any momentum, no well-defined propagator. Attempts like that of \cite{Haussling} to arrive at well-defined Green's functions by the introduction of symmetry-breaking terms lead to predictions that are gauge-dependent. Although it must be realized that in field theory it is the $S$-matrix elements rather than the Green's functions themselves that ought to be unambiguous (hence gauge-independent), and that in zero dimensions there is simply no physical room for a truncation of Green's function to $S$-matrix element, we see the result of \cite{Haussling} as a clear indication that this approach misses the essential physics of the model. Indeed, the domain on which the Goldstone modes live is a compact (hypersphere) rather than an infinite Cartesian space, so that the usual Gaussian approximation to their behaviour cannot expected to be appropriate.

\subsection{The Green's functions}

Rather than computing the path integral for the $O(N)$ model, we shall compute the (full) Green's functions of the theory. Let us define
\bqa
\Delta^{(2)}_{\al_1\al_2} &=& \delta_{\al_1\al_2},\nl
\Delta^{(4)}_{\al_1\al_2\al_3\al_4} &=& \delta_{\al_1\al_2}\Delta^{(2)}_{\al_3\al_4} + \delta_{\al_1\al_3}\Delta^{(2)}_{\al_2\al_4} + \delta_{\al_1\al_4}\Delta^{(2)}_{\al_2\al_3},\nl
\Delta^{(6)}_{\al_1\al_2\al_3\al_4\al_5\al_6} &=& \delta_{\al_1\al_2}\Delta^{(4)}_{\al_3\al_4\al_5\al_6} + \delta_{\al_1\al_3}\Delta^{(4)}_{\al_2\al_4\al_5\al_6} + \delta_{\al_1\al_4}\Delta^{(4)}_{\al_2\al_3\al_5\al_6} \nl
& & \delta_{\al_1\al_5}\Delta^{(4)}_{\al_2\al_3\al_4\al_6} + \delta_{\al_1\al_6}\Delta^{(4)}_{\al_2\al_3\al_4\al_5},
\eqa
and so on, where the indices $\al$ take values in $1,2,\ldots,N$, and $\delta$ is the Kronecker symbol. We have immediately the properties
\bq
\delta^{\al_1\al_2}\delta^{\al_3\al_4}\cdots \delta^{\al_{2n-1}\al_{2n}} \Delta^{(2n)}_{\al_1\cdots\al_{2n}} = {2^n\Gamma(n+N/2)\over\Gamma(N/2)},
\eq
and
\bq
\Delta^{(2n)}_{\al\al\cdots\al} = {(2n)!\over n!\;2^n} \quad \textrm{(no summation over $\alpha$)}.
\eq
By the $O(N)$ symmetry of the theory, we have that the expectation value of an odd number of fields vanishes, while
\bq
\left\langle
\f_{\al_1}\f_{\al_2}\cdots\f_{\al_{2n}}\right\rangle = \Delta^{(2n)}_{\al_1\cdots\al_{2n}}{\Gamma(N/2)\over 2^n\Gamma(n+N/2)} {R_{2n}\over R_0},
\eq
where, disregarding $n$-independent constants,
\bqa
R_{2n} &\propto& \int d^N\vvf\;(\vvf^2)^n\; \exp \left( -{1\over\hbar} \left[ {\lambda\over24}(\vvf^2)^2 - {1\over2}\mu\vvf^2 + {3\over2} {\mu^2\over\lambda} \right] \right) \nl
&\propto& v^{2n} \int d^N\vec{f} (\vec{f}^2)^n \exp \left( - {1\over4g} (\vec{f}^2-1)^2 \right) \nl
&\propto& v^{2n} \intl_{-1}^\infty ds\;(s+1)^{n-1+N/2}\exp\left(-{s^2\over4g}\right)\nl
&\approx& v^{2n} \suml_{m\ge0}{g^m\Gamma(n+N/2) \over m!\Gamma(n-2m+N/2)}.
\eqa
Here $\vec{f}$ and $g$ are defined the same as in the $Z_2$-model calculation, see (\ref{dimlessvars}). In the last step we have approximated by including the $\vec{f}^2-1\equiv s<-1$ part of the integral. The error made here involves terms that are not just non-perturbative but truly negligibly small if $g$ ($\hbar$) is small. To leading order in $g$, we therefore have $R_{2n}=R_0$, and for even $N$, $R_{2n}$ is a finite polynomial in $g$. The full Green's functions of this $O(N)$ model are therefore given by
\bqa
\langle\f_{\al_1}\f_{\al_2}\cdots\f_{\al_{2n}}\rangle &=& v^{2n} {1\over2^n}\Delta^{(2n)}_{\al_1\cdots\al_{2n}} \left[\suml_{m\ge0}{g^m\over m!\Gamma(n-2m+N/2)}\right] \left[\suml_{m\ge0}{g^m\over m!\Gamma(-2m+N/2)}\right]^{-1} \nl
&=& v^{2n} \Delta^{(2n)}_{\al_1\cdots\al_{2n}}{\Gamma(N/2)\over 2^n\Gamma(n+N/2)} + {\cal O}(g).
\eqa

\subsection{The $\f_1$ sector}

To compute the complete effective action it is usefull first to look at the $\f_1$ sector alone. Let us therefore take all the external-field indices $\al$ in the previous paragraph to be 1. We find, then, the full Green's functions to leading order:
\bq
\langle\f_1^{2n+1}\rangle = 0, \quad \langle\f_1^{2n}\rangle = {v^{2n}(2n)!\Gamma(N/2)\over4^nn!\Gamma(n+N/2)},
\eq
for $n\ge0$. From the set of full Green's functions we can compute the connected Green's functions through
\bq
\phi_1(J) = \sum_{n \geq 0} \frac{1}{n!} \left( \frac{J}{\hbar} \right)^n \langle \f_1^{n+1} \rangle_{\mathrm{conn}} = \hbar \frac{Z'_1}{Z_1} 
\eq
with
\bqa
Z_1(J) &=& \suml_{n\ge0}\;{1\over n!}\left({J\over\hbar}\right)^n \left\langle{\f_1^n}\right\rangle = \suml_{n\ge0}\;{1\over n!}\left({vJ\over2\hbar}\right)^{2n} {\Gamma(N/2)\over\Gamma(n+N/2)}\nl
&=& \Gamma(N/2)\left({vJ\over2\hbar}\right)^{1-N/2} I_{N/2-1}\left({vJ\over\hbar}\right), \label{Z1bessel}
\eqa
where $I$ denotes the modified Bessel function of the first kind. Note that the object $Z_1$ is just a device to go from the full Green's functions to the connected ones, it has no direct interpretation as a path integral.

Since, to leading order in $\hbar$, $Z$ is a function of the combination $vJ/\hbar$, so is $\phi_1/v$, and therefore $J$ is $\hbar/v$ times a function of $\phi_1/v$. Straightforward expansion techniques allow us to arrive at $J$, the derivative of the effective action, as a function of $\phi_1/v$:
\bqa
J &=& {\hbar\over v} \;\suml_{k\ge1} \; \left( {\phi_1\over v} \right)^{2k-1}N^kc_k, \nl
c_1 &=& 1, \quad c_2\;\;=\;\;{1\over(N+2)}, \nl
c_3 &=& {N+8\over(N+2)^2(N+4)}, \quad c_4\;\;=\;\;{N^2+14N+120\over(N+2)^3(N+4)(N+6)}, \nl
c_5 &=& {(N+24)(N^3+6N^2+112N+448)\over(N+2)^4(N+4)^2(N+6)(N+8)}, \nl
c_6 &=& {N^5+40N^4+892N^3+2016N^2+57216N+322560\over (N+2)^5(N+4)^2(N+6)(N+8)(N+10)},
\eqa
and so on. For $N=1$, we recover the `flat' result of the previous section (Of course $O(1)$ is no symmetry group, $N=1$ simply means we consider a theory with \emph{one} field, the above is still valid for this case.):
\bq
J\rfloor_{N=1} = {\hbar\over2v}\ln\left({1+{\phi_1\over v}\over1-{\phi_1\over v}}\right),
\eq
whereas, at the outer extreme, we have for $N\to\infty$:
\bq
\lim_{N\to\infty}\;J = {N\hbar\over v}{{\phi_1\over v}\over1-\left({\phi_1\over v}\right)^2}. \label{ninfty}
\eq
This last limit makes sense as long as $N\to\infty$ with $N\hbar$ bounded. It can easily be checked that, if $N\hbar$ is indeed bounded, the limiting form (\ref{ninfty}) also follows directly from the asymptotic form of the Bessel function for large order as given in \cite{Stegun}.

\section{The effective action in the flat region}
We shall now extend the above results to the full $O(N)$-theory. Because of the presence of a continuous manifold of minima, perturbation theory around a single minimum, and the corresponding diagrammatic techniques, are inadequate to compute the Green's functions. However, as soon as the {\em connected\/} Green's functions have been established, it is a simple manner to derive from them the (formally) 1PI amplitudes of the theory. Although these are not diagrammatically computable, they themselves can nevertheless serve as effective vertices in an effective diagrammatic approach, in which the connected two-point functions serve as propagators. 

Now, consider the effective action of the $O(N)$ model. The effective action must depend on $\vvf$ only via $\vvf^2$. In an effective diagrammatic expansion, this implies that if an effective (tree!) diagram contains a certain $\f_k$ as an internal line, that $\f_k$ must also occur as at least 2 external lines. Therefore, the effective diagrams contributing to $\langle\f_1^{2n}\rangle$ can only contain internal lines of type $\f_1$, and hence the effective vertices involved only contain $\f_1$ legs. Therefore,  the part of the effective action $\Gamma$ that contains only (powers of) $\f_1$ is {\em identical\/} to the effective action one can calculate from the $J$ computed above. We therefore have
\bq
\Gamma(\vvf) = \left\lfloor\int J(\phi_1)\;d\phi_1 \right\rfloor_{\phi_1^2\to\vvf^2}.
\eq
For general $N\ge2$, we therefore have
\bq \label{effactON}
\Gamma(\vvf^2) = \hbar\suml_{k\ge1} {N^kc_k\over2k}\left(\left({\vvf\over v}\right)^2\right)^k.
\eq
As a special case,
\bq
\lim_{N\to\infty}\Gamma(\vvf^2) = -{N\hbar\over2}\ln\left(1-\left({\vvf\over v}\right)^2\right).
\eq
The expressions for the effective action become undefined when $|\vvf|\approx v$: this, as we have already seen in the $N=1$ case, simply signals the transition from the `flat' region, where  $\Gamma$ is of order $\hbar$, to a `steep' region, where $\Gamma$ is of order unity. However, this is not directly visible from (\ref{effactON}), to see this one really has to sum the series. (\ref{effactON}) Is only usefull for $|\vvf| \ll v$. 

It is however easy to derive a formula giving the flat solution around $|\vvf|=v$. For very large $J$ we have from (\ref{Z1bessel}):
\bqa
Z_1(J) &\propto& \left( {vJ\over2\hbar} \right)^{1-N/2} I_{N/2-1} \left( {vJ\over\hbar} \right) \nl
&\sim& {1\over\sqrt{2\pi}} \left( {vJ\over2\hbar} \right)^{1-N/2} \frac{e^{{vJ\over\hbar}}}{\sqrt{{vJ\over\hbar}}} \quad \Rightarrow \nl
\phi_1(J) &=& \hbar {Z'_1\over Z_1} \sim \hbar \left[ \left(1-{N\over2}\right) {1\over J} + {v\over\hbar} - {1\over2J} \right] = \hbar \left[ {1-N\over2} {1\over J} + {v\over\hbar} \right].
\eqa
Now we see that when $J$ goes to $+\infty$, $\phi_1$ approaches $1$ from below, so that it was indeed appropriate to take the asymptotic expansion for the Bessel function. The last relation is easily inverted and integrated to give:
\bq
\Gamma = -\hbar {N-1\over2} \ln \left( 1-{\phi_1\over v} \right)
\eq
for $N \geq 2$. We get the complete effective action again by replacing $|\phi_1|$ by $|\vvf|$:
\bq
\Gamma = -\hbar {N-1\over2} \ln \left( 1-{|\vvf|\over v} \right).
\eq

In the two pictures below one can see the derivative of the flat solution (this derivative contains all physical information) for two values of $N$. Both the approximation for small $|\vvf|$ (We included all terms up to order $|\vvf|^{28}$ in (\ref{effactON}).) and the approximation for $|\vvf|$ around $v$ are plotted.
\begin{center}
\includegraphics[width=8cm, angle=-90]{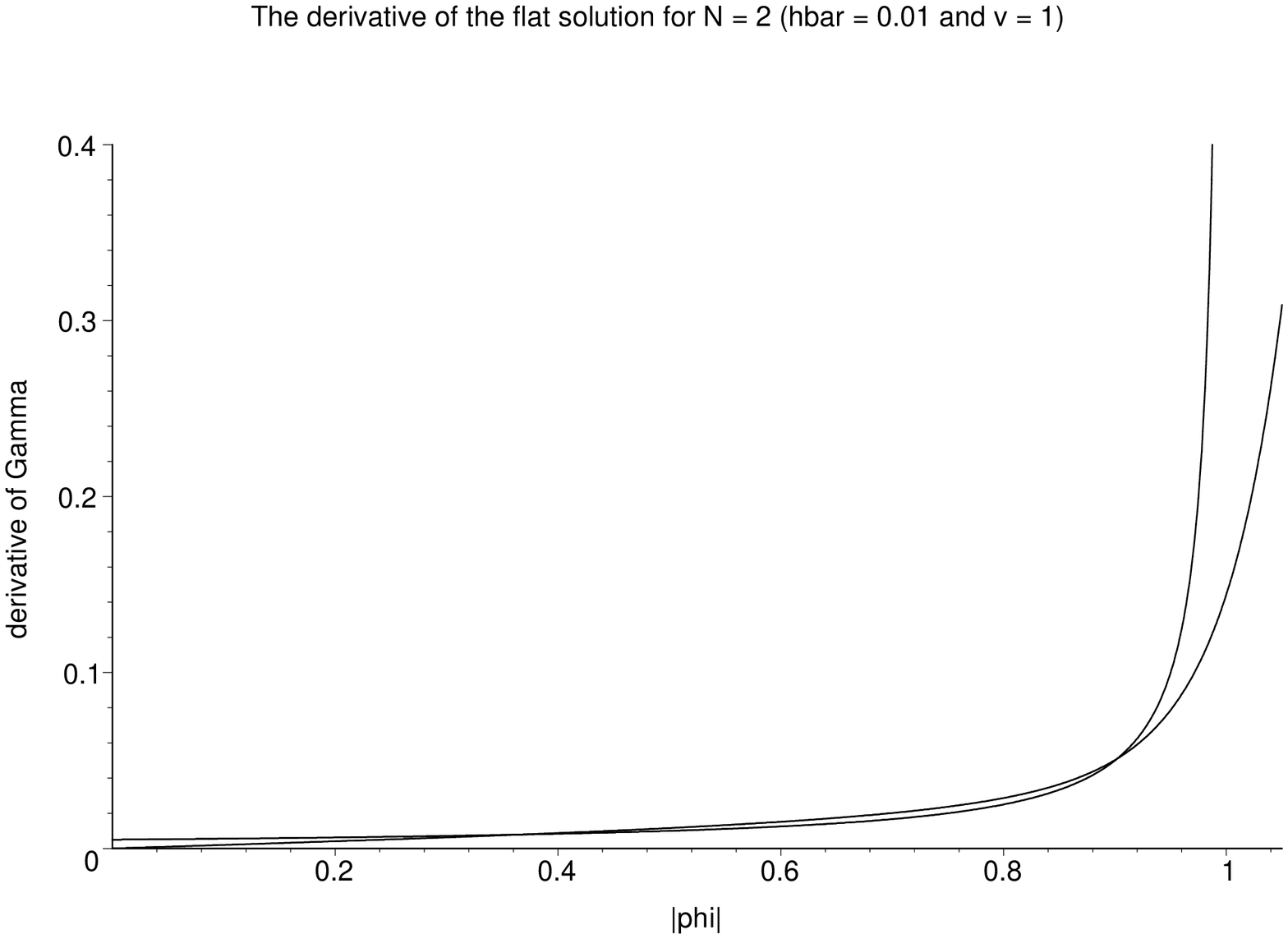} 

\vspace{20pt}

\includegraphics[width=8cm, angle=-90]{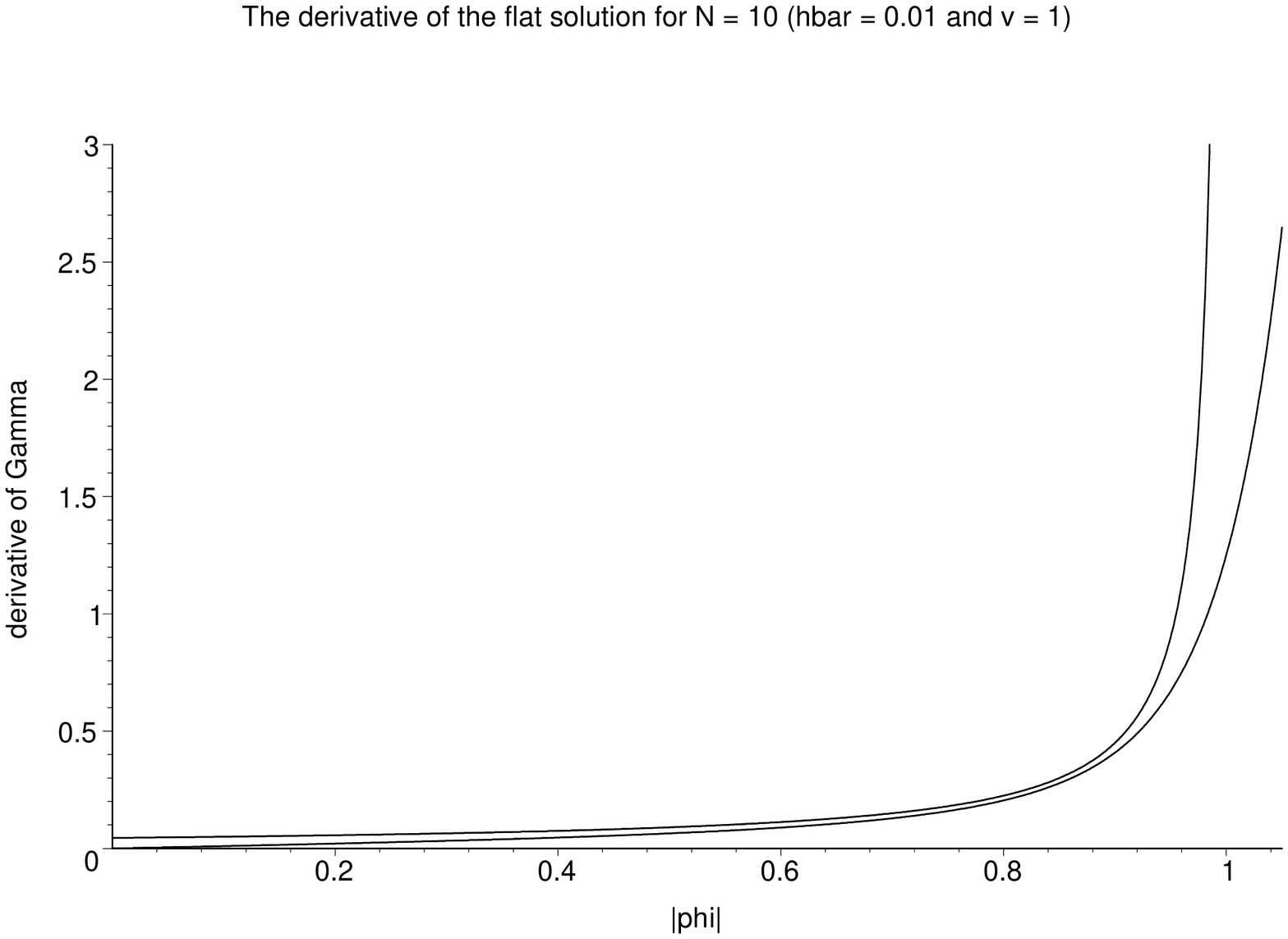}
\end{center}
Notice that the approximation to the flat solution around $|\vvf|=v$ does not go through the origin, whereas the approximation around $|\vvf|=0$ does.

We see that around $|\vvf|=0$, where the flat solution is a very good approximation to the true effective action, the effective action has derivative zero. Our effective action is concave (at least at lowest order) and we have \emph{no} spontaneous symmetry breaking.

This means our results do not agree with the result in \cite{Haussling}, where, for $N=2$, the effective action at leading order is found to be equal to the bare action.

The reader may wonder why we have not simply evaluated the path integral for our action through the Schwinger-Dyson equations, but used the more roundabout derivation using the $\Delta$'s defined in the beginning of this section. The reason is the fact that to leading order in $\hbar$ the expectation values $R_{2n}$ are all equal. This is true for any action with an $O(N)$-symmetric set of minima at $|\vvf|=v$. Therefore our result for the effective action (\ref{effactON}) is correct to leading order for a much larger class of bare actions, for instance $S \sim (\vvf^2-v^2)^4$, for which the Schwinger-Dyson equation is a seventh order differential equation.

\section{Conclusion}

We have performed an explicit computation of the effective action
for zero-dimensional Euclidean theories in which the bare action exhibits
several (or infinitely many) minima. In all cases the effective action
is well-defined for all field values, as well as purely concave, in contrast
to the bare action. We have given arguments against the validity of
the na\"ive perturbative symmetry-breaking approach. In fact, as already
discussed by \cite{Haussling}, the perturbative approach will unavoidably
lead to massless Goldstone bosons, for which the propagator is undefined
in zero dimensions. The attempt in \cite{Haussling} to repair this situation
by including symmetry-breaking terms and ghost fields in the action leads to
Green's functions that are gauge-dependent. By itself this is not a problem,
although in zero dimensions it is not possible to truncate Green's functions
into $S$-matrix elements so as to inspect {\em their\/} gauge-(in)dependence;
but we view the occurrence of a massless Goldstone boson as the unwanted effect
of treating the tangential degree of freedom in the $SO(2)$ model, which 
is defined on a compact space (in this case the circle) as a massless free
field defined on the tangent space to the symmetry breaking point ({\it i.e.},
the infinite straight line). Indeed, the more careful treatment reveals that,
around the origin of the space of field values, 
we end up with two fields of the $SO(2)$
symmetric theory actually both having the same mass, of order
${\cal O}(\hbar)$, rather than one of them being massless and the other one
more massive, with a mass of ${\cal O}(1)$.

The absence of spontaneous symmetry breaking in zero dimensions is not a
surprise: indeed in {\it e.g.\/} \cite{Coleman} it was already discussed that
spontaneous symmetry breaking is absent in fewer than three dimensions.
Our emphasis is, rather, on the form of the effective action itself, since
this is generally viewed as being a better approximation to the `actual
physics' than the bare action. In all cases (and, as we shall show, also
in one dimension \cite{NextPaper}) the effective action follows, for small
$\hbar$, closely the bare action in case the source $J$ is large, of order
unity: this is the `steep' solution. For small values of the source $J$,
of order ${\cal O}(\hbar)$, the effective action is still perfectly concave, 
and does
{\em not\/} approximate the bare action. If the bare action is described by
a `Mexican hat', the effective action has the shape of a shallow bowl, with the
minimum at the origin. In this region, the response of the field to a small
change in the source is extremely drastic, but in the absence of sources the 
symmetry is not broken.

A final remark is in order here. It might be thought that, if the effective
action approximates the bare one for all source values except for those
of order ${\cal O}(\hbar)$, this is a very small range of limited interest.
In fact, it is the other way around: if we see our field $\varphi$ as a
Higgs field coupling to, say, a fermion field $\psi$, the coupling term 
in the action of this larger theory would read  $g\varphi\bar{\psi}\psi$ with
a coupling constant $g$ of order unity. Since the natural size of a field
fluctuation such as $\psi$ is of order ${\cal O}(\sqrt{\hbar})$, this
coupling terms acts as a source term for the Higgs field of precisely 
size ${\cal O}(\hbar)$. The `weak source' region is therefore physically
extremely relevant.

\section{Acknowledgements}

This research was supported by E.U. contract numbers HPRN-CT-2000-00149 and HPMD-CT-2001-00105. Also we acknowledge Costas Papadopoulos for very useful discussions.

\end{document}